\documentclass[showpacs]{revtex4-1}
\usepackage{graphicx}
\usepackage{amsmath}
\usepackage{latexsym}
\usepackage{subfigure}
\usepackage{color}
\newcommand{\sigmael}{\sigma^{\textrm{n}}}

\newcommand{\zetabar}{\overline{\zeta}}
\newcommand{\kbar}{\overline{k}}
\newcommand{\alphabar}{\overline{\alpha}}
\newcommand{\Ubar}{\overline{U}}
\newcommand{\xilbar}{\overline{\xi}_l}
\newcommand{\lambbar}{\overline{\lambda}}
\newcommand{\edd}{\end{document}}

\begin{document}

%\doublespacing
%\include{mycommands}
%\bibliography{articles1,books}
%
\title{Active gel model of amoeboid cell motility}
\author{A.~C. Callan-Jones\footnote{Email address of corresponding author:  andrew.callan-jones@univ-montp2.fr}}
\affiliation{Laboratoire Charles Coulomb,\\ UMR 5521 CNRS-UM2, Universit\'e Montpellier II,  34095 Montpellier Cedex 5, France  }
\author{R. Voituriez}
\affiliation{Laboratoire de Physique Th\'eorique et Mati\`ere Condens\'ee, UMR 7600, Universit\'e Pierre et Marie Curie/CNRS, Paris, France}
\affiliation{Laboratoire Jean Perrin, FRE 3231 CNRS /UPMC, 4 Place Jussieu, 75255
Paris Cedex}

%\date{\today}
%
%
%\setstretch{1.5}
\begin{abstract}
We develop a model of amoeboid cell motility based on active gel theory.  Modeling the motile apparatus 
of a eukaryotic cell as a confined layer of finite length of poroelastic active gel permeated by a solvent, we first show that, due to active stress and gel turnover,
 an initially static and homogeneous layer can
undergo a contractile-type instability to a polarized moving state in which the rear is enriched in gel polymer. This agrees qualitatively with motile
cells containing an actomyosin-rich uropod at their rear.  We find that the gel layer settles into a steadily moving, inhomogeneous
state at long times, sustained by a balance between contractility and filament turnover. In addition, our model predicts an optimal 
value of the gel-susbstrate adhesion leading to maximum layer speed, in agreement with cell motility assays.
The model may be relevant to motility of cells translocating in complex, confining environments that can be mimicked experimentally 
by cell migration through microchannels.  

\end{abstract}

\maketitle

%\section{Introduction}
%\begin{itemize}
%\item Precedents on active gel hydrodynamics
%\item Precedents on concentration fluctuations in polymer solutions, micellar solutions, colloidal
%solutions
%\item Modification of traditional hydrodynamics to include quasi-slow variables
%\item Importance of chemical reactions in sustaining relative flow.  
%\item Existence of viscous stresses $\sim \nabla v_p$.  References:  Doi and Onuki, J. Phys 1992; Tanaka, PRE.  
%\end{itemize}

%The mechanical properties of eukaryotic cells are strongly determined by their cytoskeleton.  
\section{Introduction}
\label{sec:Intro}
%\noindent References:
%\begin{enumerate}
%\item Pushing off the wall ~\cite{Hawkins:2009fk}
%\item Spontaneous contractility ~\cite{Hawkins:2011uq}
%\item Verkhovsky fragments ~\cite{Verkhovsky_1999}
%\item Yam et al keratocytes ~\cite{Yam:2007lr}
%\item Amoeboid motility
%\end{enumerate}

Cell motility plays a role in key physiological processes such as wound healing, morphogenesis, and immunological response.  
In the past few decades it has been a main focus of cell biology, leading to the identification of many of the molecular players involved.  In particular, 
the actin cytoskeleton, a network of polar, semiflexible protein filaments  has been shown to be an essential part of the motility machinery.  This network is highly dynamic and out-of-equilibrium: filament polymerization and depolymerization give rise to spontaneous network flows and molecular motors such as myosin-II interact with actin, behaving as active crosslinkers and exerting stresses 
on the network.  

The actin cytoskeleton has also attracted much attention in the physics community, since it represents an important example 
in the class of active living matter, that also includes, for example, bacterial suspensions, microtubule-kinesin solutions, and bird flocks~\cite{Marchetti_Active_Matter}.  A defining trait of these systems is a constant source of energy input, leading to non-equilibrium behavior 
such as pattern-forming instabilities and collective motion.  The actin cytoskeleton, driven by hydrolysis of ATP, has been the focus of active gel theory~\cite{Kruse_generic,Julicher_Phys_Rep,Joanny_3fluid,CJ_gels}, a hydrodynamic approach providing a framework for quantitatively understanding cell biological phenomena such as the formation of contractile rings during cell division~\cite{Salbreux:2009ve}, cortical flows during development~\cite{Mayer:2010fk}, and, of course, cell motility~\cite{Kruse_lamell,Hawkins:2009fk,Hawkins:2011uq}. 
%

%Spatial distribution of actin cytoskeleton

%Interplay between polymerization and depolymerization and myosin-induced contractility

%Spontaneous symmetry breaking:  Yam, Verkhovsky, Poincloux

%Active stresses essential for symmetry breaking:  Tjhung et al (theory), Yam (experiment)

%Motility requires actin turnover:  treadmilling in polar cytoskeleton, or bulk turnover \\
%\begin{itemize}
%\item 2D crawling: mostly treadmilling dependent
%\item 3D motion: depend on contractility  (Sixt~\cite{Sixt:2012vn}, Tjhung~\cite{Tjhung:2012kx})
%\end{itemize}

%Motility resulting from a linear instability of the gel not mediated by a regulatory species (see Hawkins, Bois)

%Two-fluid models accounts for differences between polymer and center of mass velocities

%Importance of external forces on motility (Yam? Verkhovsky? others???)

%Maintain cell polarity despite delocalization of main polarity modules (see Sixt JCB) (~\cite{Poincloux_3D_motility},~\cite{Hawkins:2011uq}, Petrie et al~\cite{Petrie:2012ys})

%One-dimensional models:  Ziebert et al., Bayly et al., 
%Two-dimensional models: Kruse et al.

Work on cell motility has focused on two basic migration modes.  In the first mode, sometimes referred to as the mesenchymal mode, cell crawling on a flat surface is powered by the lamellipodium, a thin, fan-shaped structure in front of the cell body that is dense with actomyosin~\cite{Pollard_Borisy}.  Crawling motility on 2D surfaces is understood as follows:  polymerization of actin at the lamellipodium leading edge along with anchoring of new filaments to the substrate via focal adhesions generates a pushing force against the cell membrane, while contractility of the actomyosin network at the rear pulls the cell body forward.  Steady moving states are maintained by turnover of actin filaments, a process known as treadmilling.  Polarization of the actomyosin gel, strong adhesion, and high cell shape anisotropy of the lamellipodium are characteristic of this type of motility~\cite{Keren:2008zr}.  Theoretical modeling based on one-component active gel theory has successfully accounted for several features of this migration mode, including the dependence of cytoskeletal flows on contractility and cell speed on filament turnover~\cite{Kruse_lamell}.

However, the environment that a cell usually encounters \textit{in vivo} is a 3D extracellular matrix, providing only weak attachment points that the cell can 
push off of, thus making the lamellipodium mode not well suited.  In fact, several recent studies have shown that cell types such as fibroblasts~\cite{Petrie:2012ys}, leukocytes~\cite{Lammermann:2008vn}, and cancerous human breast cells~\cite{Poincloux_3D_motility}, can migrate in complex 3D geometries via an amoeboid mode.  In addition, tumor cells can undergo a mesenchymal-to-amoeboid transition in the presence of protease inhibitors that prevent extracellular matrix degradation, thus favoring amoeboid-type movement which enables the cells to squeeze through small spaces~\cite{Wolf_2003,Sahai:2003bh}.  In the amoeboid mode, there is often no leading edge actin polymerization, and motility is strongly dependent on myosin II-driven contractility of the actin network, generating cytosolic fluxes and often producing blebs.  Despite growing interest in amoeboid motility, there has been far less theoretical investigation done into this mode than into its lamellipodial cousin.  Contractility-induced permeation of the cytosol through the actin cytoskeleton is a critical aspect of this migration mode, and a theoretical description requires treating the cytoskeleton as a multi-component, poroelastic gel.  This approach has successfully been used to quantitatively describe blebbing~\cite{Charras:2005bh}.  We propose here a model of amoeboid cell motility based on multi-component active gel theory~\cite{CJ_gels}.

Precisely how cell polarity and directional movement are maintained remain open questions.  Polarity in 2D crawling cells is thought to be sustained by a complex signaling network involving localization of Rho family proteins 
to the lamellipodium leading edge~\cite{Sixt:2012vn}.  In contrast, establishment of polarity in cells executing amoeboid motility is less well understood.  For instance, polarity-associated signaling molecules are delocalized during 3D fibroblast migration~\cite{Petrie:2012ys}.  
Yet, a recent model of round breast tumor cells migrating in extracellular matrix~\cite{Poincloux_3D_motility} has uncovered a means 
for the cell to migrate persistently without a preexisting leading edge:  myosin II-induced contractility generates an instability of the homogeneous actin cortex of a stationary cell, breaking the spherical symmetry, and giving rise to cortical actin flows and cell motion~\cite{Hawkins:2011uq}.  
This results confirms the critical role played by contractility in the spontaneous symmetry breaking of static round keratocytes~\cite{Yam:2007lr}, keratocyte fragments~\cite{Verkhovsky_1999}, and 
active droplets~\cite{Tjhung:2012kx}.  Nonetheless, it is not yet clear how the two fundamental processes involved in cell motility, namely, active contractility and turnover of the actin cytoskeleton, relate to the persistence and speed of cell motion.  
 
%, and predicts that underlying instabilities of the actomyosin gel can provide a signaling-free pathway for cell polarity and motion. 
In this paper, we develop a model of amoeboid cell motility, treating the cell as a confined layer of a multi-component, poroelastic, active gel in a one dimensional geometry that is  relevant to motility assays of cells confined in microchannels~\cite{Hawkins:2009fk,Jacobelli:2010kx}.  We first elucidate the linear instabilities of a homogeneous, stationary gel, which results in a polarized, moving state.  We find self-propelled moving steady-states at long times,
and study the relation between motility, contractility, adhesion to the substrate, and filament turnover.  

Several recent theoretical studies have found pattern-forming instabilities in active gels of infinite extent~\cite{Liverpool_2003,Ziebert:2004qf,Voituriez_phasediag,Banerjee_Marchetti,Bois_pattern}.  By considering a finite film geometry,
we see how these instabilities, coupled frictionally to a substrate, yield center of mass motion.  Cell motility is a vastly intricate problem, involving feedback between biochemical signaling and mechanics.  However, by following the modeling approach developed here, based on conservation laws and force balance, we are able to uncover basic dependencies that necessarily remain in more complex models.  

\section{Model of amoeboid cell motility}
\label{sec:GelModel}
\begin{figure}[h]
\begin{center}
\includegraphics[width=13cm,trim=0mm 0mm  50mm 20mm,clip=true]{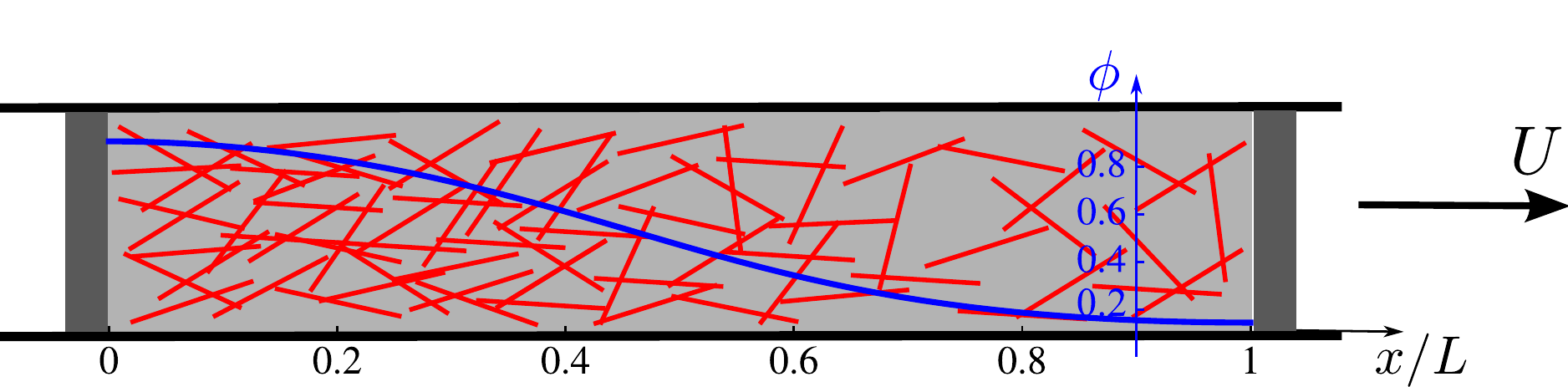}%3cm
\caption{A layer of moving active gel of length $L$. The gel consists of polymer filaments (red lines) permeated by a solvent (light gray); it is bound above and below by two fixed horizontal surfaces, and on the left and right by two moveable pistons.  The blue curve, determined by numerically solving Eq.~\eqref{eq:ContinuityPhi_NonDim}, indicates the polymer volume fraction in the steady-state of the layer moving towards the right, and illustrates that the layer is enriched in polymer at its rear.  The parameter values used in the steady-state solution of Eq.~\eqref{eq:ContinuityPhi_NonDim} are $\phi_0=0.5$, $\zetabar=33$, $\kbar_d=200$, $\alphabar=100$, and $\xilbar=0.46$.  The corresponding steady state speed of the film is $\Ubar_{\infty}=16.7$.}
\label{figs:MovingLayerSchematic}
\end{center}
\end{figure}

%\subsection{Model overview}
%\label{subsec:overview}
Eukaryotic cell motility, whether lamellipodium-based crawling of keratocytes on a 2D surface~\cite{Pollard_Borisy} or translocation of quasi-spherical cells through a 3D extracellular matrix~\cite{Poincloux_3D_motility}, depends on polymerization and depolymerization of the actin filaments and on active stresses within the actin network.  Here, we model confined cell motility as follows.  Consider a layer of an isotropic polymer network in a solvent held between two flat surfaces.  The layer is initially homogeneous and at rest with respect to the surfaces.  The length of the layer is $L$ and its height is $h$.  To represent loads, such as the nucleus trailing the motile apparatus or a viscous load encountered by the cell anterior, we introduce two identical ``pistons" at each end of the gel layer; see Fig.~\ref{figs:MovingLayerSchematic}.  This geometry is in fact relevant to motility assays of cells confined in microchannels~\cite{Hawkins:2009fk,Jacobelli:2010kx}.

\subsection{Active gel description}
\label{subsec:activegeldescrip}
Active gel theory provides a continuum framework for studying the long time and length scale flows of the 
cytoskeleton~\cite{Kruse_generic,Joanny_3fluid,CJ_gels}.  In this theory, the cytoskeleton is described as an actin polymer network permeated by a cytosol.  The network is cross-linked by several types of proteins such as $\alpha$-actinin and filamin; rheological studies have shown that actin networks undergo solid-to-liquid behavior on timescales greater than 1-10 s~\cite{Wottawah_2005}.  Moreover, actin filaments interact with molecular motors such as myosin II, that, via ATP hydrolysis, exert 
active stresses on the network.  

\subsubsection{Polymer mass balance}
In our model, gel motion arises from active stress-induced gel phase separation and network turnover.  To see how this comes about we start with a simplified description of active gel theory, in which spatial dependence occurs only along the $x$-direction. First, the polymer volume fraction $\phi(x,t)$, whose value is $\phi_0$ in the homogeneous state, satisfies the continuity equation
\begin{align}
%\frac{d\rho}{dt}+\rho\partial_x v&=0
%\label{eq:ContinuityRho}\\
\partial_t\phi-U\partial_x\phi+\partial_x J_p=-k_d(\phi-\phi_0)\,.
\label{eq:ContinuityPhi}
\end{align}
In this equation, $U(t)$ and $J_p(x,t)$ are, respectively, the layer velocity and polymer flux with respect to the fixed frame and $k_d$ is the bulk filament depolymerization rate.  In our description, filaments undergo uniform bulk polymerization with the rate $k_d\,\phi_0$, assuming that diffusion of free monomers is fast enough that we may consider their concentration to be fixed at its homogeneous, unperturbed value.   Our approach is different from those that restrict filament polymerization to one or more boundaries, thereby breaking the symmetry of the problem by hand.  We note that the coordinate $x$ is measured with respect to a fixed point in the frame moving with velocity $U$.  Incompressibility of the combined polymer plus solvent system implies that the length of the layer is constant. 

\subsubsection{Forces on the gel}
Flows involved in cell motility occur at low Reynolds numbers and thus the gel layer is always at mechanical equilibrium.  The forces acting on the polymer component of the gel include a force due to stresses in the network; an osmotic force due to composition inhomogeneities; polymer-solvent friction; and polymer-substrate friction.  The polymer-solvent friction force is much smaller than the polymer-substrate force: the ratio of the two is
$\eta_s/(\xi \ell^2)$, where $\eta_s$ is the solvent viscosity, $\xi$ is the polymer-substrate friction coefficient, and $\ell$ is the network mesh size.  Taking $\eta_s=10^{-3}$ Pa.s, $\xi=10^{16}$ Pa.s/m$^2$~\cite{Kruse_lamell} (for a layer thickness of 1 $\mu$m), and $\ell=50$ nm, we find $\eta_s/(\xi \ell^2)\sim 10^{-4}$, and thus we may neglect momentum exchange between polymer and solvent.  As a result, the balance of forces on the polymer component of the gel gives
\begin{equation}
\partial_x\left(\sigmael-\Pi\right)=\xi J_p\,,
\label{eq:PolymerForceBalance}
\end{equation}
where $\sigmael$ is the $xx$-component of the network stress and $\Pi$ is the osmotic pressure.  
We note that in Eq.~\eqref{eq:PolymerForceBalance} the polymer-substrate friction force is written as $\xi J_p=\xi\, \phi\, v_p$, as opposed to the usual form $\xi\,v_p$, where $v_p$ is the polymer velocity, to account for the dependence of the friction force on polymer volume fraction.  

%\subsubsection{Elastic stress}
The network stress, $\sigmael$, contains passive and active contributions.  Here, we consider the liquid limit of the gel, valid on timescales greater than the 
Maxwell time $\tau_{\textrm{M}}=1-10$ s.  In this limit, the passive part of the network stress, arising from polymer convection and crosslink
remodeling, has the viscous form $\eta_p \partial_x J_p$, where $\eta_p$ is the polymer viscosity~\cite{Doi_Onuki,Milner_1995,CJ_gels}.  However, this contribution does not qualitatively affect the flows and motile behavior of the gel layer and we do not consider it further~\cite{Note_on_viscous_stress}.  The active part, on the other hand, is essential for motility. Since the magnitude of the active stress, arising from motor activity on filaments, increases with $\phi$ a simple choice for the network stress is~\cite{note_on_active_stress}
\begin{equation}
\sigmael=\zeta\phi\,.
\label{eq:ElasticStress}
\end{equation}
A linear dependence of the active stress on polymer concentration has also been considered in recent work on motile active droplets~\cite{Tjhung:2012kx}.  We consider positive values of the activity coefficient $\zeta$, corresponding to contractile behavior of active actin networks~\cite{Bendix:2008fk,Koenderink:2009zr}.  Moreover, $\zeta>0$ can give rise to contractile instabilities and gel phase separation~\cite{Bendix:2008fk,Banerjee_Marchetti}.  We also note that the active stress has been interpreted in the context of blebbing and amoeboid motility as a myosin-induced active hydrostatic pressure~\cite{Charras:2005bh,Charras:2008ly}.  Though this active stress results from the isotropic part of the stress tensor~\cite{CJ_gels}, it is physically meaningful 
since the polymer component of the gel is generally compressible.  

The osmotic pressure, $\Pi$, is formally given by $\Pi=-F/V+\phi\,\delta F/\delta\phi$, where $F(\phi,\partial_x\phi)$ is the $\phi$-dependent part of 
the gel free energy and $V$ is the gel volume~\cite{Milner_1995}.  Physically, $\Pi$ acts to saturate the linear instability
causing gel phase separation and to smooth the interface between polymer-rich and polymer-poor regions.  Thus, a simple, phenomenological
form for $\Pi$ is
\begin{equation}
\Pi=\alpha (\phi-\phi_0)^3-\gamma \partial_x^2\phi\,,
\label{eq:OsmoticPressure}
\end{equation}
where $\alpha$ and $\gamma$ are positive coefficients~\cite{note_on_osmotic_pressure}.  A term linear in $\phi$ in $\Pi$, describing filament diffusion, has been omitted since it can be absorbed 
into $\sigmael$.  Finally, combining Eqs.~\eqref{eq:ContinuityPhi}-\eqref{eq:OsmoticPressure} we obtain the equation of motion for $\phi$
\begin{align}
\partial_t\phi-U\partial_x\phi+\frac{1}{\xi}\partial_x^2(\zeta\phi-\alpha(\phi-\phi_0)^3+\gamma\partial_x^2\phi)=-k_d(\phi-\phi_0)\,.
\label{eq:ContinuityPhi2}
\end{align}
Equation~\eqref{eq:ContinuityPhi2} has, minus the convective term $-U\partial_x\phi$, the form of a Cahn-Hilliard equation including a chemical reaction term~\cite{Cahn_Hilliard}.   This model has, in particular, been used to study spinodal decomposition and steady states of binary solutions with chemical exchange between the components~\cite{Glotzer_MonteCarlo,Glotzer_1995}.
By including the convective term, we will show how Eq.~\eqref{eq:ContinuityPhi2} predicts inhomogeneous, moving steady-states.

\subsubsection{Boundary conditions and global force balance}
The solution of Eq.~\eqref{eq:ContinuityPhi2} requires the specification of four integration constants, determined from boundary conditions at the layer ends, and the unknown layer velocity
$U(t)$.  
No-flux boundary conditions on the polymer are expressed as $J_p(0,t)=U\phi(0,t)$ and $J_p(L,t)=U\phi(L,t)$.  These equations imply no-flux of the solvent at the boundaries, as well.  In addition, assuming that there are no specific surface energies or dissipation at $x=0$ and $x=L$ implies that $\partial_x\phi(0,t)=0$ and $\partial_x\phi(L,t)=0$~\cite{Burch:2009zr,Note_on_VariatBCs}. Finally, $U(t)$ can be found by requiring that the gel layer plus the end pistons are at mechanical equilibrium, namely
\begin{equation}
\xi_l U+\xi\int_0^L J_p\,dx=0\,,
\label{eq:GlobalForceBalance}
\end{equation}
where $\xi_l$ is the friction coefficient between the piston loads and the surface.

\subsection{Contractile instability, phase separation, and moving states}

\subsubsection{Linear instability}
In this section, we demonstrate that an instability of the homogeneous, stationary gel layer gives rise to gel phase separation and layer motion. We will focus on the onset of unstable states, depending on the activity parameter $\zeta$, the depolymerization rate $k_d$, and the drag coefficient $\xi_l$.  First, a straightforward linear stability analysis can be performed in the limit 
$\xi_l\to\infty$, i.e., for fixed pistons, in which case the eigenmodes $\phi(x,t)-\phi_0\sim e^{\lambda^{\infty}_n t}\cos{(n\pi x/L)}$, $n=1,2,3,\ldots$, substituted 
into Eq.~\eqref{eq:ContinuityPhi2}, yield the dispersion relation
\begin{equation}
\lambda^{\infty}_n=-k_d+\frac{1}{\xi}\left(\frac{n\pi}{L}\right)^2\left(\zeta-\left(\frac{n\pi}{L}\right)^2\gamma\right)\,.
\label{eq:DispersionRelation}
\end{equation}
Equation~\eqref{eq:DispersionRelation} reveals that activity drives the linear instability of the gel; the mixing effect of filament depolymerization suppresses long wavelength modes; and the smoothing term $-\gamma\,\partial_x^2\phi$ cuts off short wavelength modes.  
%The effects of $\zeta$ and $k_d$ on the gel stability are qualitatively unchanged when the constraint of fixed pistons 
%is relaxed.  As a result of the contractile instability and inhomogeneities in $\phi$, polymer fluxes develop and, as we will see, give rise to net motion of the layer.

%In the following, when considering a linearly unstable gel, we suppose that $\zetabar$ lies between $\pi^2$ and $4\pi^2$, so that the $n=1$ mode is the most unstable.  

\begin{figure}[h]
\begin{center}
\includegraphics[clip=true,width=10cm]{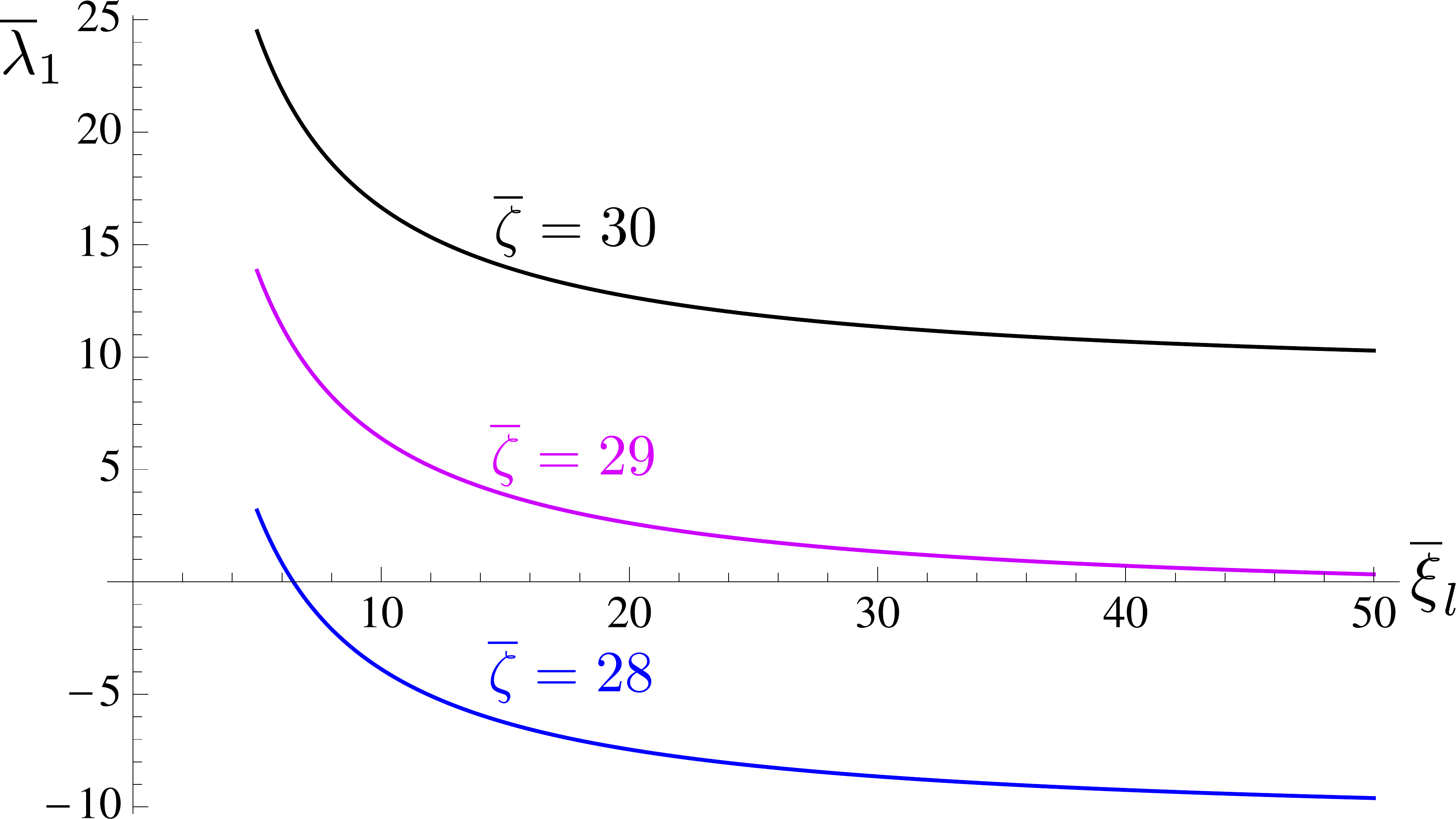}
\caption{Growth rate of $n=1$ mode versus piston drag. $\lambbar_1$ versus $\xilbar$ is shown for three different activities, as indicated.
The other parameters are $\kbar_d=190$ and $\phi_0=0.5$.}
\label{figs:lambda_vs_xil}
\end{center}
\end{figure}

The stability analysis for the case of moving pistons is more involved, as the third spatial derivative of the eigenmodes no longer vanishes at the layer ends. 
To highlight the role of $\zeta$, $k_d$, and $\xi_l$, we introduce the dimensionless quantities 
$\zetabar=\zeta L^2/\gamma$, $\kbar_d=k_d\xi L^4/\gamma$, $\xilbar=\xi_l/(\xi L)$, $\alphabar=\alpha L^2/\gamma$, $\overline{J}_p=J_p\xi L^3 /\gamma$, and $\Ubar=U \xi L^3/\gamma$.  Furthermore, we non-dimensionalize $x$ by $L$ and $t$ by $\xi L^4/\gamma$, keeping the old variable names for the new ones.  As a result, the dimensionless equation of motion for $\phi$ is
\begin{align}
\partial_t\phi-\Ubar\partial_x\phi+\partial_x^2(\zetabar\phi-\alphabar(\phi-\phi_0)^3+\partial_x^2\phi)=-\kbar_d(\phi-\phi_0)\,.
\label{eq:ContinuityPhi_NonDim}
\end{align}
Writing $\phi(x,t)-\phi_0=e^{\lambbar_n t}\phi_n(x)$, inserting this ansatz into Eq.~\eqref{eq:ContinuityPhi_NonDim}, and keeping only linear terms in $\phi_n$ yields
\begin{equation}
\phi_n''''(x)+\zetabar\phi_n''(x)=-(\kbar_d+\lambbar_n)\phi_n(x)\,,
\label{eq:CH_linear}
\end{equation}
where primes ($'$) denote differentiation with respect to $x$.  Writing the layer velocity as $\Ubar(t)=\tilde{U}_n e^{\lambbar_n t}$, the boundary conditions on the third derivatives of $\phi$ and the flux $\overline{J}_p$ at 
$x=0$ and $x=1$ read
\begin{subequations}
\label{eq:BCs}
\begin{align}
\phi_n'(0)=0,\quad &\phi_n'(1)=0\,; \label{eq:BC_Variat}\\
\phi_n'''(0)=\tilde{U}_n\phi_0,\quad &\phi_n'''(1)=\tilde{U}_n\phi_0\,;\label{eq:BCs_NoFlux}
\end{align}
\end{subequations}
while the global force balance is
\begin{equation}
\xilbar \tilde{U}_n+\zetabar \phi_n\Big|_0^1+\phi_n''\Big|_0^1=0\,.
\label{eq:ForceBalance_linear}
\end{equation}
The modes with non-zero $\tilde{U}_n$ are antisymmetric about $x=1/2$. We thus obtain
\begin{equation}
\phi_n(x)=A_{n+}\sin{\left[k_{n+}(x-1/2)\right]}+A_{n-}\sin{\left[k_{n-}(x-1/2)\right]}\,,
\end{equation}
where $A_{n\pm}$ are constants and 
\begin{equation}
k_{n\pm}=\sqrt{\frac{\zetabar}{2}\pm \frac{1}{2}\sqrt{\zetabar^2-4(\kbar_d+\lambbar_n)}}\,.
\end{equation}
The condition for non-trivial values of $A_{n\pm}$ is obtained by setting the determinant of the two-by-two matrix obtained from the boundary and force balance conditions,
Eqs.~\eqref{eq:BCs}-\eqref{eq:ForceBalance_linear}, to zero. 
%
%\begin{equation}
%\left|
% \begin{array}{cc}
%k_{n+}^3 \cos{(k_{n+} /2)} & k_{n-}^3 \cos{(k_{n-}/2)} \\
%\frac{k_{n+}\xilbar\zetabar}{\phi_0}\cos{(k_{n+}/2)}+2(\zetabar^2-k_{n+}^2)\sin{(k_{n+}/2)} \quad
%&
%\frac{k_{n-}\xilbar\zetabar}{\phi_0}\cos{(k_{n-}/2)}+2(\zetabar^2-k_{n-}^2)\sin{(k_{n-}/2)} 
%\end{array} 
%\right|=0\,.
%\end{equation}
%
This results in the characteristic equation 
\begin{align}
 & k_{n+} \cos{(k_{n+}/2)} \left(\frac{k_{n-}^3\xilbar}{\phi_0}\cos{(k_{n-}/2)}-2(\zetabar^2-k_{n-}^2)\sin{(k_{n-}/2)} \right) \nonumber \\
&\quad\quad {}-
k_{n-} \cos{(k_{n-}/2)} \left(\frac{k_{n+}^3\xilbar}{\phi_0}\cos{(k_{n+}/2)}-2(\zetabar^2-k_{n+}^2)\sin{(k_{n+}/2)} \right)=0\,,
\label{eq:CharacEq}
\end{align}
which can be solved numerically for the growth rate $\lambbar_n$ for given $\zetabar$, $\kbar_d$, and $\xilbar$. In the following, we will focus on values of $\zetabar$ for which $n=1$ is the most unstable mode, corresponding, approximately, to $\pi^2<\zetabar<9\pi^2$.  However, in principle higher, odd modes are excited for larger $\zetabar$, though the lowest odd mode is likely the most relevant for cell motility.
We note that a finite drag coefficient $\xi_l$ affects  the dispersion relation in a non-trivial way, as can be seen from Eq.~\eqref{eq:CharacEq}. 
Qualitatively, we expect that the smaller the piston drag, the more unstable the gel layer is.  Indeed,
Fig.~\ref{figs:lambda_vs_xil} reveals that the growth rate of the first mode, $\lambbar_1$, decreases as 
$\xilbar$ increases.  
%This can be understood by noting that as $\xilbar$ decreases, the slope of $\phi$ increases
%at the boundaries, resulting in a shorter mode wavelength and thus a larger $\lambbar_1$.  
Interesting, if, in the polymer force balance, the polymer-solvent drag were to dominate the polymer friction with the substrate, i.e., $J_p-\phi\,U\sim \partial_x(\sigma^{\textrm{n}}-\Pi)$, then $\lambbar$ would no longer depend on $\xilbar$, since
no-flux at $x=0,1$ would mean that the eigenmodes for $\phi(x,t)-\phi_0$ are always of the form $\cos{(n\pi x)}$.

\begin{figure}[h]
\begin{center}
\includegraphics[clip=true,width=14cm]{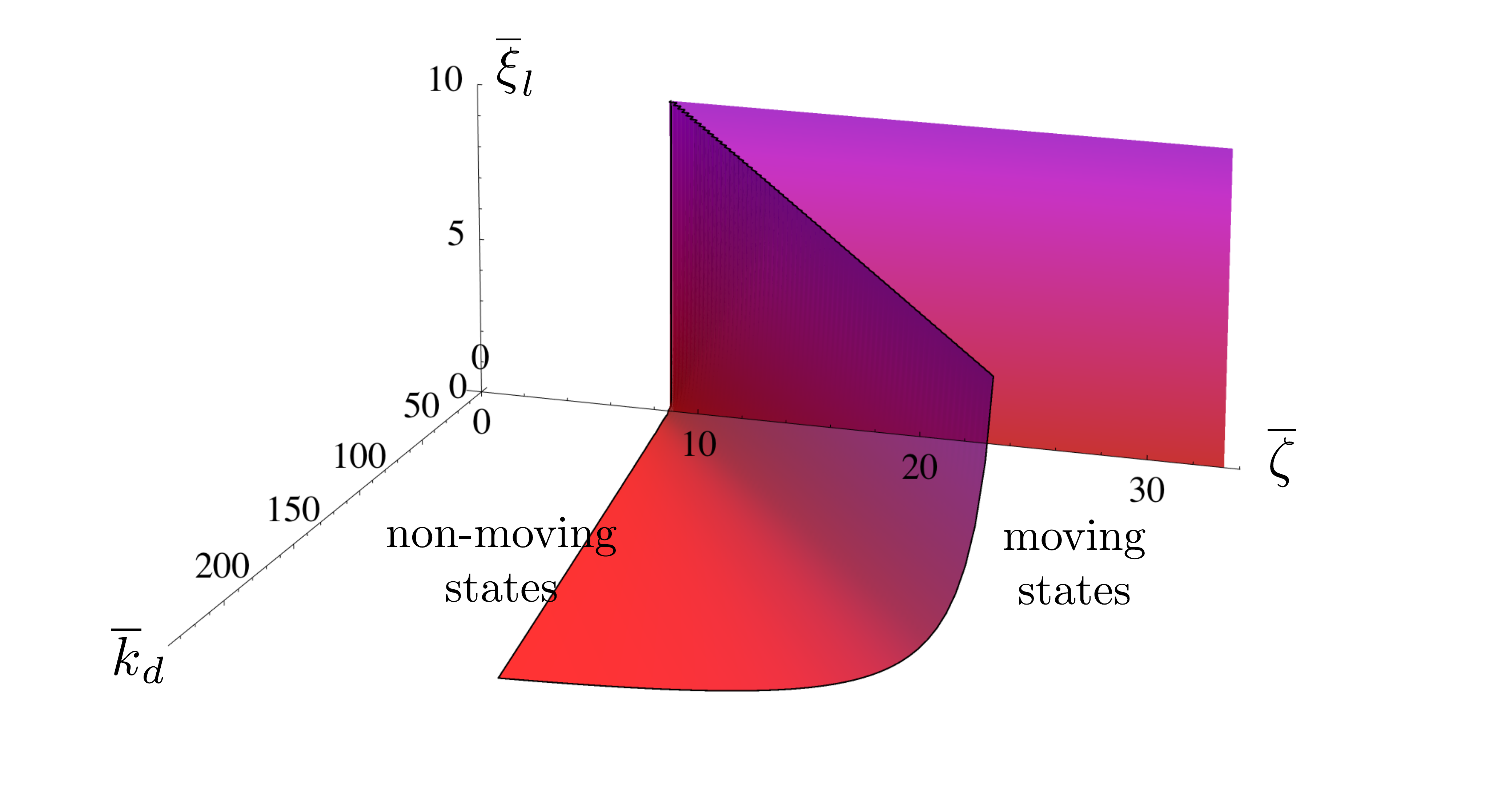}%{xim_vs_Da_modified.eps}
\caption{
State diagram in the $(\kbar_d,\zetabar,\xilbar)$ space. Moving states exist in the region bound by the surfaces $\lambbar_1=0$ and $\kbar_d=0$.  The initial polymer volume fraction is $\phi_0=0.5$.
}
\label{figs:StateDiagram}
\end{center}
\end{figure}

\subsubsection{Gel phase separation and moving steady-states}

In the regime $\lambbar_1>0$, a contractile instability of the gel layer appears, which is triggered by active stress, but is counteracted by filament turnover.  However, filament turnover is essential for moving steady-states of the gel to exist.  Indeed, if $\kbar_d=0$, polymer conservation implies that 
at steady-state $\overline{J}_p(x)=\Ubar\phi(x)$; substituting this into the global force balance, Eq.~\eqref{eq:GlobalForceBalance}, and solving for $\Ubar$, it follows that the layer speed is zero.  

An inhomogeneous polymer density profile and filament turnover give rise to moving states in the regime $\lambbar_1>0$. The state diagram in Fig.~\ref{figs:StateDiagram} indicates the region in the $(\zetabar,\,\kbar_d,\,\xilbar)$ parameter space in which such moving steady-states exist, bound by the surfaces $\kbar_d=0$ and $\lambbar_1=0$. The moving states are characterized by an inhomogeneous profile $\phi(x)$  and a non-zero steady state velocity $\Ubar_{\infty}$, which are obtained by numerically solving the non-linear Eq.~\eqref{eq:ContinuityPhi_NonDim}, completed by boundary conditions and global force balance. Figure~\ref{figs:MovingLayerSchematic} shows an example of a moving steady-state for non-zero $\kbar_d$ that develops after the initial instability of the gel, and in which the layer is enriched in polymer at its rear.  The polymer volume fraction, $\phi$,  is indicated by the blue curve in Fig.~\ref{figs:MovingLayerSchematic}.  
%Note that, according to the no-flux conditions, the slope of $\phi$ at the layer ends is proportional to the layer speed $\Ubar$.  
For non-zero $\kbar_d$, net filament depolymerization occurs in parts of the layer where $\phi>\phi_0$, whereas net polymerization occurs in regions where $\phi<\phi_0$.  As a result, polymer flux is directed from right to left in Fig.~\ref{figs:MovingLayerSchematic}; exchange of momentum between the gel and the substrate provides a rightward force and drives motion of the layer to the right.

%Supposing that the initial, homogeneous, non-moving state of the gel is unstable ($\lambbar_1>0$), we could in principle calculate the layer velocity $\Ubar(t)$ for a given piston friction $\xilbar$ for each $t$, by solving the force balance Eq.~\eqref{eq:GlobalForceBalance}.  However, for numerical purposes,  it is simpler to pose the problem differently: for a given imposed layer speed $\Ubar(t)$, starting from zero and tending to a constant value $\Ubar_{\infty}$, what is the value of $\xilbar$ such that Eq.~\eqref{eq:GlobalForceBalance} holds at steady-state?  This procedure allowed us to calculate numerically the steady state layer speed $\Ubar_{\infty}$.
%

\begin{figure}[ht]
%\centering
\subfigure[]{
\hspace{0.5cm}
\includegraphics[width=0.45\textwidth,trim=0mm 0mm 0mm 0mm,clip=true]{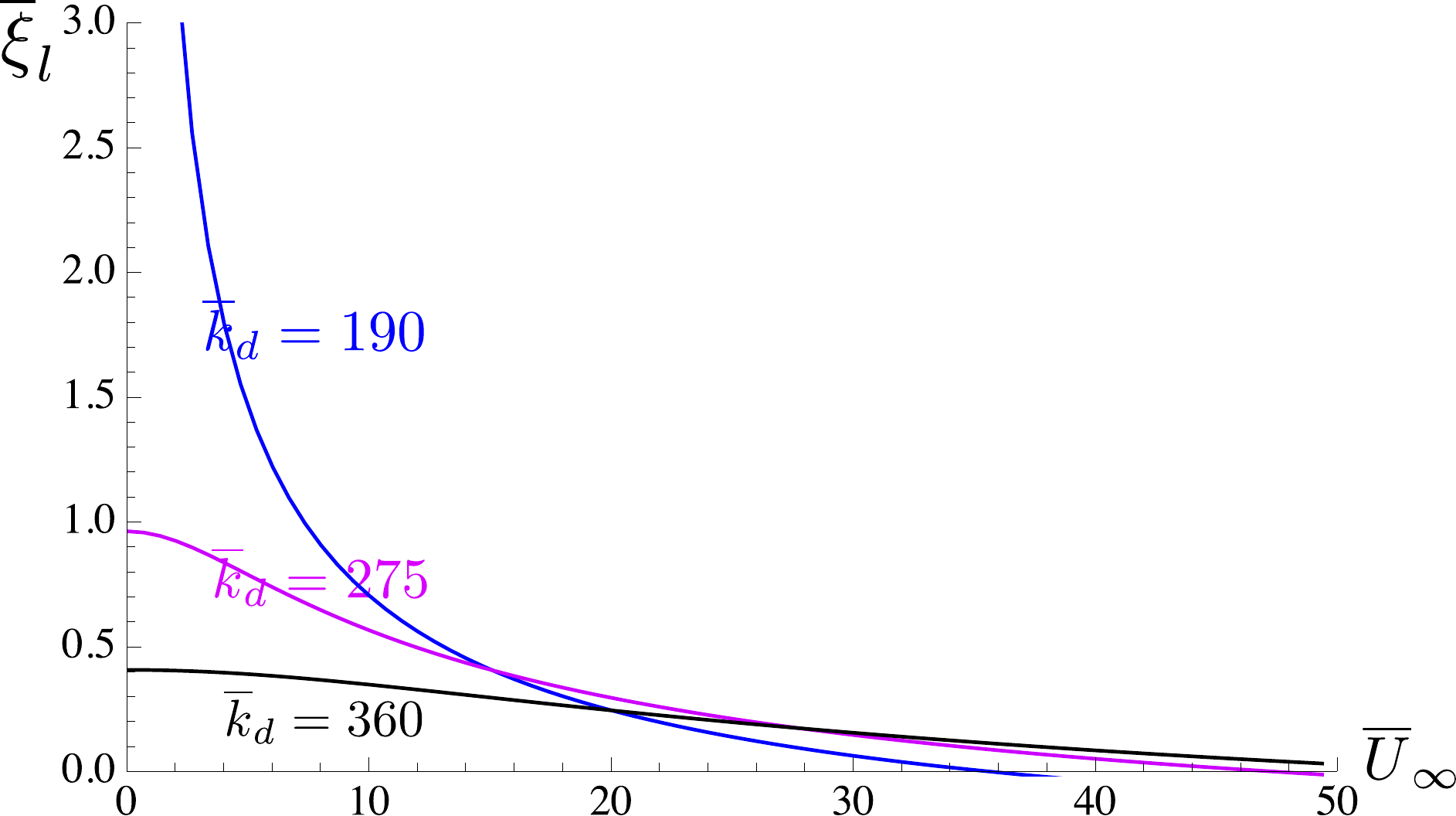}
\label{fig:xim_vs_U_vary_kd}
}
%\hspace{2cm}
\subfigure[]{
\includegraphics[width=0.45\textwidth,trim=0mm 0mm 0mm 0mm, clip=true]{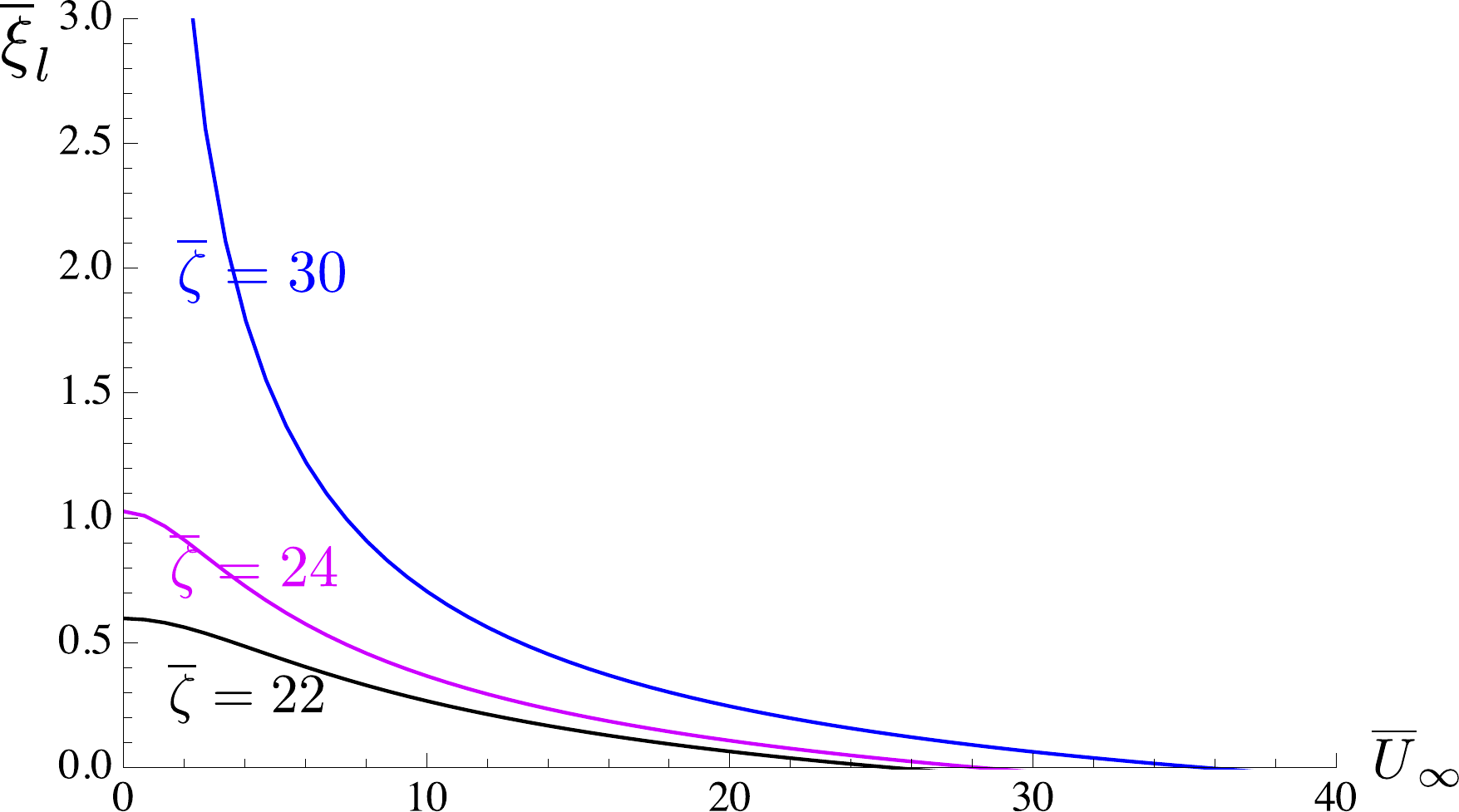}
\label{fig:xim_vs_U_vary_zeta}
}
\caption{\label{fig:xim_vs_U} 
Load friction coefficient $\xilbar$ versus steady-state layer speed $\Ubar_{\infty}$.  (a) $\xilbar$ vs $\Ubar_{\infty}$ is 
shown for three indicated values of $\kbar_d$ and $\zetabar=30$. (b) $\xilbar$ vs $\Ubar_{\infty}$ is 
shown for three indicated values of $\zetabar$ and $\kbar_d=190$.  $\xilbar$ tends to a finite value as $\Ubar_{\infty}\to 0$ if 
the linear growth rate $\overline{\lambda}^{\infty}_1<0$. In (a) and (b), the parameter values used in obtaining the steady-state solution of Eq.~\eqref{eq:ContinuityPhi_NonDim} are $\phi_0=0.5$ and $\alphabar=100$.
}
\end{figure}

%Moving steady-states of the active gel layer, corresponding to static solutions of Eq.~\eqref{eq:ContinuityPhi_NonDim} with layer velocity
%$\Ubar_{\infty}$, develop at long times after the initial instability discussed above.  
The moving states can be characterized by the load force-velocity relationship for the moving layer, described here by the dependence of $\xilbar$ on  $\Ubar_{\infty}$; see Fig.~\ref{fig:xim_vs_U}. 
Figures~\ref{fig:xim_vs_U_vary_kd} and \ref{fig:xim_vs_U_vary_zeta} show that $\xilbar$ decreases with increasing $\Ubar_{\infty}$ for, respectively, fixed $\zetabar$ 
and varying $\kbar_d$ and fixed $\kbar_d$ and varying $\zetabar$.  We note that if the dispersion relation is such that  $\overline{\lambda}^{\infty}_1>0$ (obtained for $\xilbar\to \infty$), then one must have 
$\xilbar\to \infty$ as $\Ubar_{\infty}\to 0$, since only rigidly fixed pistons can prevent a linearly unstable gel from being set into motion.  On the other hand, if $\overline{\lambda}^{\infty}_1<0$ (obtained for $\xilbar\to \infty$), moving steady-state exist for $\xilbar<\xilbar^*$, where $\xilbar^*$ depends 
on $\zetabar$ and $\kbar_d$.  In fact, $\xilbar^*$ is the load friction for which the slab is marginally stable: $\overline{\lambda}_1(\zetabar,\kbar_d,\xilbar^*)=0$. The state diagram in Fig.~\ref{figs:StateDiagram} shows that
if the depolymerization rate is $\overline{k}_d$ is large enough, the mixing effect 
of filament turnover suppresses gel phase separation and self-sustained moving states no longer exist.  This illustrates the subtle role played 
by filament turnover in our model: it is necessary for moving states to be sustained, yet if it is too high, cell polarity, needed for motion, is abolished.  

%between the surfaces $\xilbar=\xilbar^*(\zetabar,\kbar_d)$ and $\kbar_d=0$.  Moving steady-states, whether initiated by gel instability or by forcing, are qualitatively similar 
%to one another, characterized by a relative enrichment of polymer network at the rear.
%Physically, states that 
%develop from a linearly stable gel subject to initial forcing arise as a result of the inhomogeneous no-flux boundary conditions, which force gel phase separation 
%and self-sustained motion at long times.  This idea provides insight into the forced movement of initially circular, stationary keratocyte fragments that become anisotropic and enriched in actin in their rear in the motile state~\cite{Verkhovsky_1999}.

%

\begin{figure}[h]
\begin{center}
\includegraphics[clip=true,width=8cm]{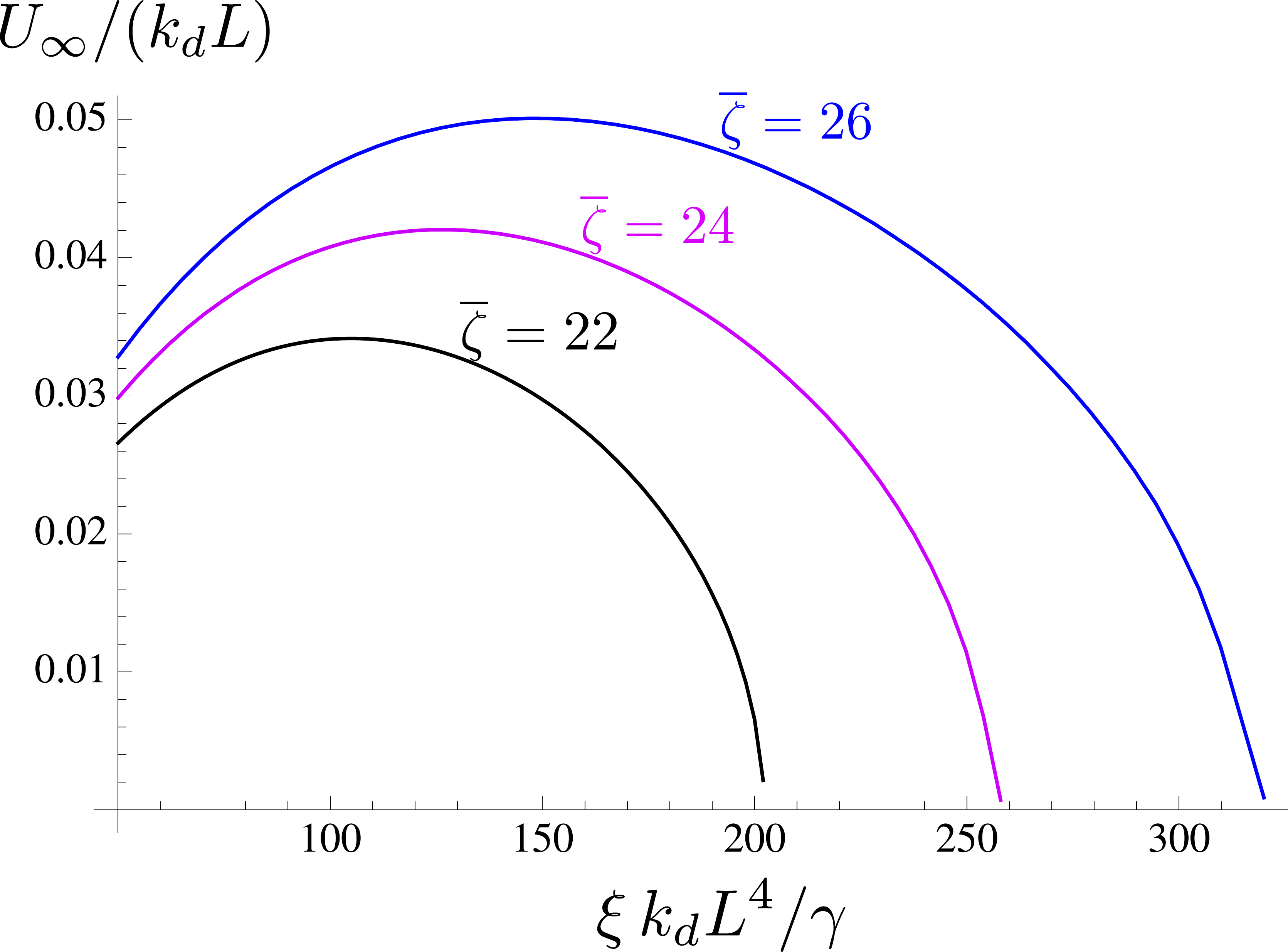}
\caption{Steady-state layer speed as a function of polymer-substrate friction.  The dimensionless speed $U_{\infty}/(k_d L)$ versus dimensionless
friction coefficient $\xi\,k_d L^4/\gamma$ is shown for three values of activity, as indicated.  The other parameters are
$\xi_l/(\xi L)=100$, $\alphabar=100$, and $\phi_0=0.5$.
 }
\label{figs:U_vs_xi}
\end{center}
\end{figure}

Motion of the gel layer requires a transfer of momentum from the substrate to the gel.  The dependence of the steady-state layer speed, $U_{\infty}$, on the polymer-substrate friction coefficient, $\xi$, allows a comparison between our model and cell motility experiments that probe the dependence of migration speed on surface adhesion.  Figure~\ref{figs:U_vs_xi} reveals that there is an optimum value of $\xi$, which maximizes $U_{\infty}$.  This prediction agrees with biphasic behaviour of migration speed as a function of surface adhesion observed in lamellipodium-driven crawling cell motility on flat surfaces~\cite{Palecek:1997fk,Gupton:2006uq} and in amoeboid motility of cells confined in microchannels~\cite{Jacobelli:2010kx}.  Our model predicts that the optimum value of $\xi$ shifts to higher values as the activity $\zeta$ increases, in agreement with the work of 
Ref.~\cite{Gupton:2006uq}.  In the model presented here, the layer speed is small for low $\xi$ because the polymers do not get enough
traction from the substrate, and is small again for high $\xi$, since, according to Eq.~\eqref{eq:ContinuityPhi2}, the amplitude of gel phase separation, and hence the layer speed, decreases with decreasing $\zeta/\xi$.  

\section{Conclusion}

We have developed a model of amoeboid cell motility that accounts for observed characteristics of this migration phenotype such as increasing cell speed with increasing contractility and a non-monotonic dependence of cell speed on surface adhesion.  The simple approach used here puts into quantitative relief the important roles played by contractility and filament turnover, considered as the two driving agents of cell motility.  
While we found the dependence of migration speed on active stress to be consistent with experiments on 3D motile cells subject to varying motor activity, our predictions involving the filament turnover rate are more nuanced.  Theoretical work has underlined the importance of the actin depolymerization rate on speed~\cite{Kruse_lamell,Zajac:2008bh}, yet our work suggests that this rate may be tuned to maintain a balance between filament recycling, needed for motion, and cell polarity, involving actin density variations.  

Amoeboid cell motility involves cytosolic permeation through the actin cytoskeleton generated by molecular motor action.  In addition,
the polymer density in the gel is non-uniform, as the cell rear is generally enriched in actin filaments.  A full accounting of these two features 
requires a multi-component active gel approach, which we have pursued here.  We note that related, multi-component models have been developed in other contexts~\cite{Kuusela:2009zr,Kimpton:2012ly}.  
Our model, treating the cytoskeleton as an active isotropic gel undergoing delocalized filament turnover, is relevant to bulk and cortical cytoskeletal flows involved in amoeboid movement.  There is, at this stage, no evidence of long range orientational
order associated with amoeboid migration, which might be related to the absence of leading edge actin polymerization in this phenotype
~\cite{Parri:2010ys,Bergert:2012vn}.  This is very different from the situation encountered in lamellipodium-based cell motility, in which leading edge polymerization sets a preferred filament ordering direction, perpendicular to the leading edge~\cite{Verkhovsky:2003qf}.     
The motility mechanism that we describe here also has similarities with the one proposed in~\cite{Hawkins:2011uq}, which was based on contractile instabilities of the actomyosin cortex generating steady state cortical flows; our present work, in fact, extends these results to cell types where the actomyosin system is not localized on the membrane but distributed in the cell bulk. We believe that such mechanisms based on actomyosin contraction localized at the back of the cell and filament recycling could in fact be very general in the context of cell motility in confinement, which can be now routinely studied in vitro in microchannel assays~\cite{Hawkins:2009fk,Jacobelli:2010kx}.

We point out finally that the modeling approach developed here may be extended to describe blebbing cell motility, a subclass of amoeboid motility involving coordinated blistering of the cell membrane and motion of the cell~\cite{Charras:2008ly}.   Cortical contraction and squeezing flow of the cytosol are implicated in this type of motion. An additional ingredient, which we have not considered here, is the convection of dissolved motors and
actin monomers by the cytosol into the advancing bleb.  Further work is needed to understand how blebbing, cytosolic flow, and protein transport 
conspire to produce cell motion.

%\section{Effect of polymer viscosity}
%\section{Coarsening behavior}
%This section is very preliminary, and just a check that numerical solution gives expected coarsening for the Cahn-Hilliard model.  Take $U=0$.  
%\begin{itemize}
%\item Take $\gamma=0.016$.  Then, $\lambda_2>\lambda_1>0$.  The initial amplitude is chosen $\phi(x)=0.01 \cos{(\pi x)}+0.01\cos{(2\pi x)}$.  The instability initially selects the $n=2$ mode, but then at long time the pattern coarsens. Rapha\"el:  see attached
%file "coarsening\_CH.mov" of $\varphi(x,t)$ vs $x$ for increasing $t$.  
%%\item Suppose the initial amplitude is $\phi(x)=\cos{(2\pi x)}$, $\gamma=0.03$.  Then $\lambda_2<0$.  However, if $1\lesssim Da \lesssim 3$, enough
%%coarsening takes places that the $n=2$ mode becomes a sort of $n=1$ mode, and an instability occurs.  
%\end{itemize}

%\bibliography{articles2,books}

%%\bibliography{articles,books}

\begin{thebibliography}{50}
\bibitem[{not({\natexlab{a}})}]{note_on_active_stress}
\bibinfo{note}{We expect the active stress to be linear in the polymer
  concentration for two reasons. First, this stress should depend on the number
  of filament crossings per unit volume. In a semi-flexible polymer network (as
  opposed to a gas of point-like particles), the number of crossings scales
  with the filament concentration. Second, if one were to assume that the
  active stress scales as $\phi^2$, then expanding about the homogeneous base
  state, $\phi=\phi_0$, results in
  $\phi^2=\phi_0^2+2\phi_0\,\delta\phi+\delta\phi^2$. Thus, the linear
  stability analysis is unchanged whether the active stress scales as $\phi$ or
  $\phi^2$. In addition, quadratic terms in $\delta\phi$ are not essential in
  producing a nonlinear, stationary state~\cite{note_on_osmotic_pressure}.}
\bibitem[{not({\natexlab{b}})}]{note_on_osmotic_pressure}
\bibinfo{note}{Alternatively, one can calculate $\Pi$ from the mixing free
  energy density, using Flory-Huggins theory. The mixing free energy is
  $f(\phi)=k_BT\rho\left[\frac{\phi}{N}\ln{\phi}+(1-\phi)\ln{(1-\phi)}\right]$,
  where, for clarity, we have omitted the interfacial, penalty term. Here,
  $\rho$ is the number of monomers per unit volume, and $N$ is the number of
  monomers per filament. Using $\Pi=-f+\phi\,\delta F/\delta\phi$, where
  $F=f\,V$, and writing $\phi=\phi_0+\delta\phi$, the osmotic pressure gradient
  is, to $O(\delta\phi^3)$, $\partial_x\Pi=k_B
  T\rho\left[\left(\frac{1}{N}+\frac{\phi_0}{1-\phi_0}\right)+\frac{\delta\phi%
}{\left(1-\phi_0\right)^2}+\frac{\delta\phi^2}{(1-\phi_0)^3}\right]\partial_x
  \delta\phi$. Suppose that a layer of active gel undergoes phase separation,
  with $\delta\phi(x,t)$ a decreasing function of $x$ between $x=0$ and $x=L$;
  $\delta\phi>0$ for $x$ near 0 and $\delta\phi<0$ for $x$ near $L$. Then, the
  terms in the above equation that have a stabilizing effect on the polymer
  flux $J_p\sim -\partial_x\Pi$, are those of order $\delta\phi$ and
  $\delta\phi^3$; the contribution of the term of order $\delta\phi^2$ to $J_p$
  is negative in regions where $\delta\phi<0$, thus reinforcing the
  instability. Thus, to capture the stabilizing effects of nonlinear terms in
  the osmotic pressure, we are free to keep only the cubic term in
  $\phi-\phi_0$, as we have done in Eq.~\eqref{eq:OsmoticPressure}.}
\bibitem[{Not()}]{Note_on_VariatBCs}
\bibinfo{note}{The choice of boundary conditions on $\partial_x\phi$ can be
  justified as follows. First, assume an interfacial, gradient free energy
  penalty of the form
  $F_i=\frac{\gamma_0}{2}\int_0^{L}\left(\partial_x\phi\right)^2\,dx$, where
  $\gamma_0$ is a positive constant. Differentiating $F_i$ with respect to time
  and performing an integration by parts yields
  $\dot{F}_i=\gamma_0\,\partial_x\phi\,\dot{\phi}|_{0}^{L}-\int_0^{L}\gamma_{0%
}\,\partial_x^2\phi\,\dot{\phi}\,dx$. The first term on the right contributes
  to the entropy balance at the boundaries; the second term results in the
  interfacial contribution $\sim \partial_x^3\phi$ to the polymer flux $J_p$.
  If there are no other surface energies, then Onsager theory applied to $x=0$
  and $x=L$ results in
  $\Gamma_s^0\,\dot{\phi}(0,t)=-\gamma_0\,\partial_x\phi(0,t)$ and
  $\Gamma_s^L\,\dot{\phi}(L,t)=-\gamma_0\,\partial_x\phi(L,t)$, where
  $\Gamma_s^{0}$ and $\Gamma_s^{L}$ are surface dissipative coefficients. Thus,
  if surface dissipation can be neglected, we recover the so-called variational
  boundary conditions, $\partial_x\phi(0,t)=0$ and $\partial_x\phi(L,t)=0$.}
\end{thebibliography}
\end{document}